# Agilité de développement des SI informatisés et outils MDE : démarche pédagogique dans un cours de conception de systèmes d'information informatisés

*Pierre-André Sunier* \*
*Steve Berberat* \*

\* Institute for Management and Information Systems, HEG Arc, HES-SO // University of Applied Sciences Western Switzerland, Neuchâtel, Switzerland

## Résumé

Les règles métiers traduites manuellement en code dans les applications de gestion sont une entrave à l'agilité des entreprises. Nos étudiants en informatique de gestion en sont-ils conscients et nos cours favorisent-ils cette prise de conscience?

Nous nous appuyons sur les concepts de pilotage par les modèles (MDA, MDE) pour démontrer, cas pratiques à l'appui, que le code peut être généré par des automates pour autant que les spécifications formelles soient suffisantes et précises.

Nous avons étendu le formalisme usuel de modélisation conceptuelle des données et créé un automate de transformation; ces deux éléments permettent à nos étudiants de constater par eux-mêmes qu'un changement de règle métier ne nécessite pas forcément la réécriture de code mais simplement de modifier le modèle et générer le code à nouveau et par là de comprendre l'intérêt d'outils MDE.

**Mots-clés :**

Ingénierie pilotée par les modèles, Architecture pilotée par les modèles, Modèle Conceptuel de Données (MCD), UML, Règles métier

## Abstract

In software development, business rules implemented by hand using programming code hinder agility of companies. Are our students in information systems aware of that? Do our lessons promote this realization ?

We use model driven concepts (MDA, MDE) in order to demonstrate, with practical examples, that source code can be automatically generated as far as formal specification are sufficient and accurate.

We have extended usual representation conventions of conceptual data modeling and developed a transformer tool. This make our students observing themselves that no source code need necessarily to be rewrite when a change of business rule occurs, but just an adaptation of the model and a new run of generation of code. In this way, they finally understand more easily the benefits of MDE tools.

# 1  Introduction

Après de nombreuses années d'enseignement, de travaux d'innovation et de mandats, nous constatons que le développement de logiciels de gestion relève, souvent, plus de l'artisanat, aussi noble soit-il, que de l'industrialisation ou de l'ingénierie.

Dans notre enseignement en informatique de gestion, de niveau Bachelor, nous observons aussi une propension des étudiants à écrire du code à partir de règles métier plutôt que de passer par une phase de modélisation qui peut ou pourrait être suivie de génération automatique de tout ou partie de code.

Durant les années 1970-2000, de nombreux travaux ont traité de l'ingénierie du logiciel et ont donné naissance à des outils CASE-Computer Aided Software Engineering; ces outils comme Designer de l'éditeur Oracle permettaient de modéliser un logiciel de gestion et de générer automatiquement le code de la structure de données, de l'intégrité des données et des éléments applicatifs.

Alors que la puissance de calcul des ordinateurs a cru régulièrement et que le potentiel de génération de code à partir de modèles aurait dû gagner en puissance, à l'image de ce qui se passe avec les prévisions météorologiques, c'est l'inverse qui nous semble s'être produit:

- La plupart des outils CASE des années 1970-2000, ont disparu du paysage et n'ont pas été remplacés.
- L'initiative ou l'approche MDA de l'OMG offre une riche couverture de modélisation en amont de la production de logiciels mais, en aval, le développement est essentiellement l'affaire de programmeurs sans relation biunivoque fiable entre modèle et code.

Ce retour ou, à tout le moins, ce recours important à la programmation humaine, nous semble aller à l'encontre de la tendance actuelle visant l'agilité des systèmes d'information.

Ce constat nous a amené à nous poser la question de recherche suivante: "Comment réconcilier modèles, dans l'esprit actuel de MDA et génération de code, dans la ligne de ce que faisaient les outils CASE?" et si réconciliation il y a, comment en faire bénéficier nos étudiants?

Pour élaborer des pistes de réponses à la question de recherche, nous avons interrogé la littérature sur:

- la nécessité d'agilité dans le processus de développement de logiciels de gestion;
- l'apport des outils CASE des années 1970-2000 et les raisons de l'abandon de ces outils;
- les opportunités, en termes d'automate de transformation de modèles, offertes par l'approche MDA de l'OMG.

Nous avons focalisé notre recherche sur la modélisation des données sachant que c'est l'élément le plus stable d'un SII comme le met en évidence Satzinger et al, (2002) "Le type de données nécessaire pour diriger une entreprise change très peu avec le temps, à l'opposé des processus de collecte des données qui eux changent souvent".

À partir de cette emphase sur la modélisation des données, nous avons travaillé sur deux axes principaux; à savoir :

- Élaboration d'une extension du formalisme usuel de modélisation conceptuelle des données en nous appuyant sur UML et les profils UML; nous visons à créer une typologie de spécifications déclaratives efficientes au travers des stéréotypes, des contraintes et des valeurs marquées d'UML. Les spécifications déclaratives doivent permettre au modélisateur[1] d'intégrer une règle de gestion ou métier sans devoir recourir à une traduction sous forme de programmation.
- Conception et développement d'un automate logiciel de transformation de modèles et de génération de code de création de structure de bases de données relationnelles (SQL-DDL) et de gestion de l'intégrité des données grâce à des APIs de tables[2]. Afin de soutenir efficacement un processus de développement itératif et incrémental, concept essentiel à un développement agile, nous nous fixons de réaliser une transformation tout aussi itérative et incrémentale.

Pour l'élaboration d'une extension du formalisme de modélisation conceptuelle de données, nous avons adossé notre travail à la littérature traitant de modélisation conceptuelle des données que ce soit d'ouvrages traitant de méthodologie comme Merise et ses multiples compléments, Software Engineering, CDM d'Oracle ou d'ouvrages traitant de langages de modélisation tels que le standard de l'OMG UML.

Pour la conception et le développement de l'automate logiciel de transformation et s'agissant des références normatives et bibliographiques, nous nous sommes appuyés sur le travail de F. Camus (2012).

---

[1] Le modélisateur est le bénéficiaire final du résultat de notre recherche.

[2] Les APIS de tables sont constituées de déclencheurs et de code au sein des bases de données relationnelles; les APIs de tables offrent des services de contrôle, de traçabilité ou encore de traitement que le langage SQL-DDL ne supporte pas. Par exemple, les APIs de tables peuvent assurer la non modification d'une valeur de colonne de table équivalent à une contrainte {frozen} en UML.

## 2 Revue de littérature

La littérature que nous avons mobilisée pour effectuer nos recherches n'est pas composée uniquement d'ouvrages et d'articles récents. Elle comprend aussi des ouvrages relativement anciens, car les concepts qui guident l'analyse des systèmes d'information d'entreprise et plus particulièrement la modélisation conceptuelle des données y sont posés. Ainsi, les travaux amenant la méthode Merise, au début des années 1980, (Tardieu et al, 1986) dans le monde francophone et les travaux de P. Chen avec ER Model dans le monde anglo-saxon, en 1976, gardent tout leur sens.

Nous commençons par relater quelques articles de la presse locale récents qui soulèvent, indirectement, la problématique de la rigidité inhérente aux processus informatisés. Cette rigidité des processus informatisés peut être partiellement solutionnée par le recours à l'automatisation de production de code, à l'image de ce que permettaient les outils CASE des années 1970-2000. Nous résumerons les apports et les concepts que véhiculaient ces logiciels sur la base de plusieurs ouvrages.

Il s'ensuit une synthèse d'un travail de recherche (Camus, 2012) qui, observant la fin des outils CASE, recense leur apport en terme d'efficience du processus de développement et explore la piste de l'initiative MDA de l'OMG en recherchant des solutions logicielles implantant les concepts MDE. Malheureusement, la conclusion de F. Camus est qu'il n'y a pas de solution au remplacement des outils CASE de l'époque, ce qui nous mène à nous poser notre question de recherche.

Les ouvrages que nous passons en revue ensuite relatent les bases sur lesquelles nous nous appuyons pour parvenir à notre proposition de solution. Dans un premier temps, nous résumons les ouvrages ayant apporté les bases de la modélisation des données. Enfin, nous étayons ces bases en ajoutant le concept de pilotage par les modèles, décrit notamment par des ouvrages traitant de l'architecture pilotée par les modèles (MDA) et de l'ingénierie pilotée par les modèles (MDE).

Pour schématiser la conceptualisation de notre revue de littérature, il nous a semblé pertinent de produire une carte de concepts telle que proposée par Rowley J. et Slack F. (2004). Cette carte est présentée en Figure 1 et sera décrite au fur et à mesure de la lecture de la revue de littérature.

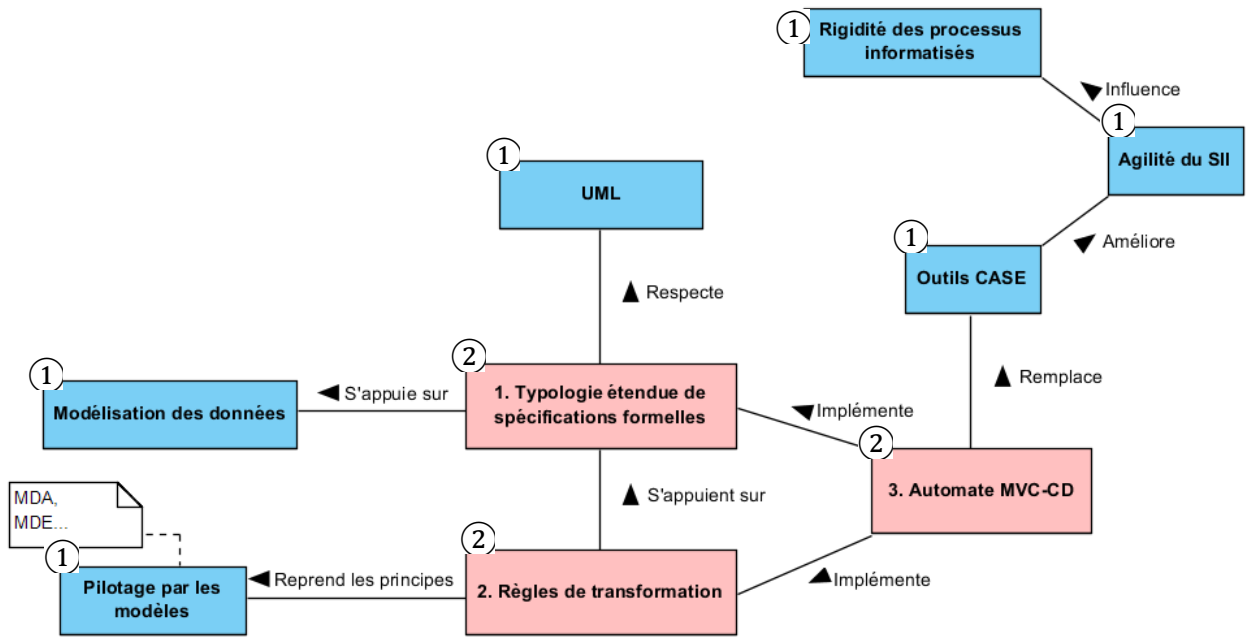

Figure 1 - Carte de concepts de notre revue de littérature

Les objets de couleur bleue ① sont les différents concepts, outils et langages existants dans la littérature et que nous avons utilisés. Ceux de couleur rouge ② présentent la contribution que nous amenons. Les relations résument l'utilisation des concepts, outils ou langages ainsi que l'influence des uns envers les autres.

Les sous-chapitres qui suivent décrivent progressivement les concepts de couleur bleue ① en exposant, pour chacun, les différents ouvrages et articles utilisés, de façon à aboutir à une présentation claire du besoin auquel tente de répondre notre contribution qui sera abordée dans le dernier sous-chapitre de la revue de littérature.

## 2.1 Rigidité des processus informatisés et agilité du SII

La presse locale relate parfois des situations où un processus métier est altéré ou même bloqué par la faute d'un logiciel. Par exemple, l'on peut citer (Kubli, 2014) qui explique se rendre à la caisse d'une chaîne de magasins en souhaitant faire valoir un cadeau de fidélité de 10% et payer une partie de ses achats avec un bon de 20 francs. La demande était tout à fait légitime et la caissière lui a confirmé le droit de cumuler ces deux avantages. La transaction n'a cependant pas pu se faire, car le système n'était simplement pas capable d'accepter les deux avantages ; le système se contentait de valider le moins avantageux pour le client.

Par cet exemple, nous remarquons bien que les processus métier informatisés peuvent être rigides. Si le processus de paiement à la caisse n'était pas informatisé, la caissière se serait contentée de réaliser le calcul à la main et le client aurait bénéficié de la réduction de 10% et du bon de 20 francs.

Un autre exemple (Theiler, 2012) montre qu'une banque suisse étudiait l'option de refuser des clients de quelques pays étrangers, alors que d'autres établissements augmentaient les frais de gestion pour ceux-ci. La raison? Elle provenait des investissements informatiques nécessaires et induits par les accords fiscaux que la Suisse établissait et, par ailleurs, continue d'établir, avec d'autres pays. Pour une banque, les coûts se chiffrent en millions de francs, ce qui rend ce type de clients peu rentables pour les petits établissements locaux.

Cette illustration montre un manque d'agilité du système d'information informatisée (SII) qui impacte de manière directe des processus métier. Bien que nous ne mettions pas en cause les performances des systèmes informatiques des banques en question, nous pouvons affirmer que, dans ce cas, plus d'agilité offrirait des solutions peut-être moins drastiques.

L'agilité, en tant qu'élément définissant la capacité à faire face à l'imprévu, est un facteur déterminant de tout développement de SII ; c'est par l'agilité que le SII pourra être ajusté aux changements organisationnels dans des délais acceptables et avec des coûts tout aussi acceptables. Des auteurs ont déjà mis cela en avant. Nous pouvons citer "Les recherches réalisées tendent à montrer que la performance des systèmes dépend surtout de leur bonne utilisation et de l'alignement des systèmes à son contexte organisationnel" (Jouirou et Kalika, 2004) ou encore "Les systèmes d'informations doivent s'adapter et évoluer pour être en adéquation avec la structure" (Baille, 2014).

Face à cette conclusion, nous nous sommes penchés sur un moyen permettant d'améliorer l'agilité du SII, à savoir: *L'utilisation d'outils d'automatisation tels que l'étaient les anciens CASE.*

## 2.2 Disparition des outils CASE

Les outils CASE étaient reconnus pour offrir des gains de productivité élevés (Banker et Kauffman, 1991). La productivité étant le rapport entre la quantité produite et les moyens mis en œuvre pour y parvenir, il est aisé de dire que l'utilisation de ce genre d'outils, pour autant que ces derniers assurent des transformations itératives et incrémentales, améliore l'agilité du SII. En effet, une meilleure productivité signifie une adaptation plus rapide du système à son organisation et explique la relation directe entre outils CASE et agilité visible en Figure 1.

Tout le monde s'accorde à dire que les outils CASE n'existent pratiquement plus aujourd'hui. Pour quelle raison ont-ils subi ce déclin? Deux hypothèses ont été avancées (Lewis, 1996). La première est leur progression sur le marché qui, bien qu'élevée, n'était pas suffisante. Selon l'auteur, "If a technology (or idea) does not achieve mainstream status quickly enough, it dies."[3]. La seconde hypothèse est "the learning rates are simply too low"[4]. C'est donc, probablement, la connaissance de ces outils et la maîtrise de leurs concepts qui ont progressé trop lentement.

---

[3] Traduction: Si une technologie ou une idée ne rejoint pas le statut grand public (populaire) assez rapidement, elle meurt.

## 2.3   Recherche d'outils productifs de conception de SII

En tant qu'enseignants de cours de conception de systèmes d'informations informatisés (SII), nous souhaitions connaître les outils existants actuellement sur le marché et qui pourraient apporter un gain en productivité comparable aux anciens CASE.

Pour obtenir réponse à notre question, nous nous sommes appuyés sur le travail de recherche de Camus (2012), qui montre une investigation faite dans le but de trouver un ou plusieurs outils qui pourraient couvrir toutes ou partie des fonctionnalités offertes par Oracle Designer, outil CASE qui n'est plus commercialisé aujourd'hui.

Les fonctionnalités recherchées durant ce travail étaient essentiellement la modélisation des données et des traitements, la transformation itérative et incrémentale de ces modèles et enfin la création et le déploiement de l'application. La recherche s'est effectuée au travers de divers types d'outils tels ceux qui se qualifient "MDE" ou encore qui sont des applications du concept "Software Factory".

Les résultats montrent qu'aucun logiciel couvrant l'ensemble des fonctionnalités souhaitées n'a été trouvé et qu'il serait nécessaire de mettre en commun plusieurs composants logiciels hétérogènes si nous souhaitions nous rapprocher d'une solution comparable.

Nous avons également interrogé la littérature qui nous apprend que MDE est utilisé par les développeurs lorsqu'ils remarquent d'eux-mêmes des fragments de code similaires qu'ils peuvent généraliser pour ensuite écrire un générateur qui se base sur un DSL (Whittle et al., 2014). Il existe ainsi un certain nombre d'outils MDE spécifiques à un domaine d'application. L'on peut citer, par exemple, une recherche portant sur la création d'un DSL pour la domotique (Jiménez et al., 2009) ou encore la création d'un outil de gestion de la traçabilité de la nourriture réalisée à partir d'un concept qui consiste à reprendre les principes MDA tout en y intégrant des avantages du Software Factory tel que l'utilisation d'un DSL (García-Díaz et al., 2010).

Ces outils peuvent pertinemment illustrer le concept MDE ; nous ne pouvons cependant pas les utiliser dans nos cours de conception de SII dans la mesure où ils sont spécifiques à un domaine et non applicables à la modélisation des données et des traitements de logiciels de gestion.

Des recherches portant sur des outils MDE propres au domaine de la conception de logiciels amènent de réels atouts. Citons par exemple ATL qui permet de spécifier, à l'aide d'un DSL s'appuyant sur OCL, des transformations de modèle à modèle, et qui peut être accompagné d'outils construits sur un environnement Eclipse (Jouault et al., 2008). Ces recherches aboutissent généralement à des composants logiciels qui, bien que fonctionnels, deviennent inaccessibles pour les étudiants de nos cours de conception de SII dans la mesure où la maîtrise des concepts inhérents à leur utilisation s'avère difficile à leur niveau d'étude.

---

[4] Traduction: Les taux d'apprentissage ont progressé trop lentement.

En s'appuyant sur ces conclusions ainsi que sur la recherche de Camus (2012), nous constatons que les outils de type CASE, complets tels qu'ils l'étaient à l'image d'Oracle Designer, n'ont actuellement pas de remplaçants.

Pour disposer d'un outil pédagogique complet, nous devrions ainsi mettre en relation plusieurs composants hétérogènes. Cette solution nous a semblé inappropriée pour deux raisons : l'assurance technique de mise en relation de ces divers composants n'est pas acquise et le résultat de ce travail ne serait certainement, ici également, pas suffisamment facile d'utilisation pour nos étudiants.

La solution que nous présentons, réalisée sur mesure, nous a permis de combler le manque d'outils CASE dans nos cours de conception de SII tout en apportant une piste de remplacement de ces outils.

## 2.4 Modélisation des données

Comme nous l'indiquions en introduction, les bases de la modélisation des données proviennent des travaux de P. Chen avec ER Model et du Modèle Conceptuel de Données ou MCD de la méthode Merise.

En plus des ouvrages fondateurs de la méthode Merise et de son Modèle Conceptuel de Données (Tardieu et al., 1986; Tardieu et al., 1985; Rochfeld et al., 1989), nous nous sommes appuyés sur d'autres ouvrages, que nous citerons plus en avant, présentant de multiples variantes de spécifications, de représentation interne ou externe ou encore, présentant des particularismes plus ou moins intéressants.

Avec l'arrivée du langage de modélisation UML, au tournant du siècle, un foisonnement de travaux et d'ouvrages ont traités de la modélisation conceptuelle des données sous forme de modèles de classes UML. L'ouvrage "De Merise à UML" (Kettani et al 1999) traite très en détail du mapping qu'il est possible de faire entre un MCD selon Merise et un diagramme de classes UML ; cet ouvrage a été une source d'inspiration et de confirmation de choix importante.

Après la publication de leur ouvrage de référence du langage de modélisation UML, Booch, Rumbaugh et Jacobson ont publié leur méthode de développement de logiciels UP, Processus Unifié (Rumbaugh et al., 2000). Au travers du Modèle du Domaine, UP installe la dimension statique du système d'information, le modèle de données, au cœur des développements orientés objet. Laman C. dans "UML-2 et les design patterns" insiste sur l'aspect statique du Modèle du Domaine: "Un modèle du domaine UP permet de visualiser les entités du monde réel et non des objets logiciels ayant des responsabilités" (Larman, 2005).

Pour faciliter et guider le travail des modélisateurs, des profils UML dédiés font leur apparition (Gornik, 2002 ; Molle, 2007) ; les profils UML nous permettront de formaliser notre typologie de spécifications formelles.

## 2.5 Pilotage par les modèles

### 2.5.1 Outils CASE

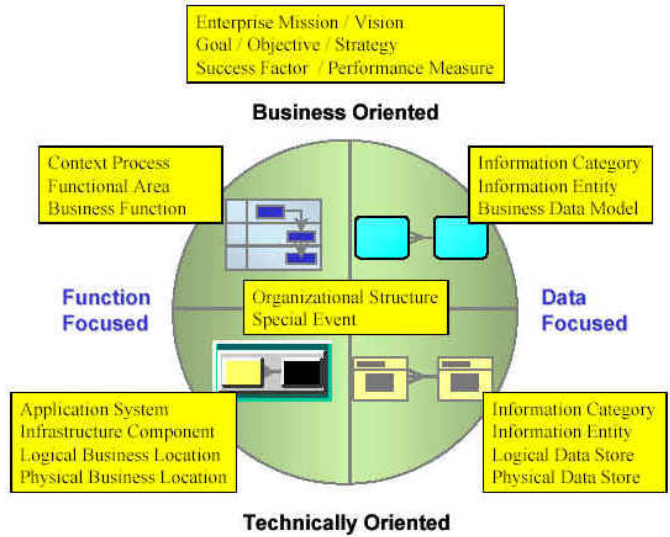

De nombreux outils CASE, dont Designer d'Oracle cité plus avant, appliquaient déjà le principe de pilotage du processus de développement de logiciels par les modèles.

Les modèles étaient de portées multiples et pouvaient être classés par niveau d'abstraction comme *Business Oriented* ou *Technically Oriented* en Figure 2.

Par contre, la terminologie de portée des modèles était très disparate et souvent propre à une méthode, à un éditeur ou à un produit.

**Figure 2 - Les modèles au centre du processus de Oracle Designer**

**[Documentation Oracle Designer]**

### 2.5.2 Architecture pilotée par les modèles

Nous nous baserons sur l'architecture MDA proposée par l'OMG pour réaliser concrètement les mécanismes de pilotage par les modèles, soit : typologie de spécifications formelles de modélisation, transformation de modèles (itérative et incrémentale) et génération de code à partir des modèles.

En plus de la spécification et du guide MDA de l'OMG, deux ouvrages nous ont aidés à comprendre les concepts généraux de MDA (Kleppe et al., 2003 ; Blanc, 2005).

MDA prône l'utilisation de modèles comme éléments de spécification. La construction d'une application passe par différentes phases.

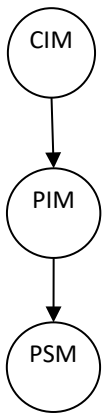

La première consiste à élaborer un ou plusieurs modèles CIM, *Computational Independant Model*. Il s'agit du modèle d'exigences qui fait abstraction de toute technologie informatique. Il est le plus pérenne et contient le vocabulaire purement métier.

Le modèle PIM, *Plateform Independant Model*, est le modèle réalisé dans la phase de conception. Il est spécifique à un concept ou une technologie informatique, mais indépendant de toute plateforme. Une transformation entre le CIM et le PIM est théoriquement réalisable et doit permettre d'établir les liens de traçabilité entre éléments des modèles (un aspect essentiel de notre transformation itérative et incrémentale).

La phase clé de MDA est la transformation du modèle PIM en un ou plusieurs modèles PSM, *Plateform Specific Model*, contenant toutes les informations nécessaires à son implémentation pour une plateforme spécifique.

La transformation de modèles PSM en code n'est pas considérée dans MDA comme une transformation de modèle. Elle s'apparente plutôt à une traduction textuelle et peut donc être facilement obtenue de façon automatique. C'est pour cela que (Kleppe et al., 2003) affirment que la génération de code "n'est pas considérée par MDA".

MDA n'impose pas UML comme langage de modélisation. Il n'impose aucun diagramme particulier pour réaliser les modèles attendus. De plus, il n'est pas une méthode de conception de SII.

### 2.5.3 Ingénierie pilotée par les modèles

Kent S. qui a réalisé la première proposition du terme MDE, affirme: "Model Driven Engineering (MDE) is wider in scope than MDA"[5] (2002). N'ayant pas les accès à l'article, nous nous sommes penchés sur l'article (Alanen et al, 2004) qui compare MDA à MDE et stipule: "MDE is a broader term that includes all models and modeling tasks needed to carry out a software project from beginning to end"[6].

La solution que nous visons a pour but de générer du code à partir de modèles à l'aide d'un outil; elle sort du cadre MDA, raison pour laquelle nous précisons ici que nous utilisons plutôt le concept MDE.

---

[5] Traduction: MDE a une portée plus large que MDA

[6] Traduction: MDE est un terme plus large qui inclut tous les modèles et les travaux de modélisation nécessaires à réaliser un projet de développement logiciel du début à la fin.

## 2.6 Notre contribution

Notre observation des pratiques actuelles du développement de logiciels de gestion et notre expérience passée des outils CASE nous incite à remettre les modèles au cœur du processus de développement pour transformer intelligemment les règles métiers en spécifications formelles et générer automatiquement le codage des dites spécifications formelles.

Ainsi, en reprenant les concepts MDA et MDE pour mettre les modèles au cœur du processus du développement de logiciels et en nous appuyant sur nos connaissances en modélisation des données, notre contribution apporte :

1. Une première typologie étendue de spécifications formelles applicables à un MCD et utilisant le langage de représentation UML.
2. Une série de règles de transformation qui décrivent comment prendre en charge ces spécifications de manière à créer des éléments physiques qui vont permettre de s'assurer de leur respect.
3. Un prototype d'automate de transformation et de génération de code qui implémente notre typologie de spécifications formelles et nos règles de transformation.

Ces trois éléments de contribution sont représentés par les objets de couleur rouge ② de notre carte de concepts visible en Figure 1.

## 3 Méthodologie

Notre recherche, le projet MVC-CD, peut probablement être qualifiée de pré-innovation dans le sens où elle tend à démontrer la faisabilité du tandem modélisation riche ou intelligente et transformation/génération de code.

De nombreux utilisateurs d'outils CASE nous ont fait part de leur frustration de la disparition de ces outils qui leur offraient des gains de productivité, en qualité et en quantité, reconnus de leurs mandants. Ces retours, empiriques, du terrain confirmaient notre propre frustration et nous ont encouragés à mener la présente recherche.

Ceci étant, notre démarche relève d'un raisonnement non formalisé en ce sens que nous avons mobilisé notre expérience :

- de la modélisation pour compiler un ensemble cohérent de règles de modélisation parmi le large éventail proposé par les chercheurs au fil du temps (Annexe A);
- des outils CASE, d'UML, de MDA, MDE et des langages de programmation pour spécifier, concevoir et réaliser un prototype d'automate de transformation/génération de code (Annexe B).

# 4   Résultats

Les bases du formalisme étendu de modélisation conceptuelle des données étant posées et l'automate opérationnel, nous présentons les résultats de notre recherche sous forme de quelques éléments pédagogiques à l'intention d'enseignants chargés de cours de conception de systèmes d'information de gestion.

Conformément, aux arguments que nous avons évoqués en introduction, nous traiterons de la modélisation conceptuelle des données dans un premier temps; ensuite, grâce à l'automate, nous montrerons comment les spécifications formelles du modèle conceptuel de données sont transformées en code exécutable garantissant l'intégrité des dites données.

## 4.1   Modèle conceptuel de données au formalisme étendu

Le modèle conceptuel de données (MCD) est réalisé sous forme de diagrammes de classe UML; les spécificités du méta-modèle "MCD" sont réalisées avec les mécanismes d'extension UML, à savoir : des stéréotypes, des valeurs marquées et des contraintes.

L'étudiant est assisté dans la réalisation d'un MCD à l'aide d'un profil UML, défini au cours du projet MVC-CD. Seuls les stéréotypes et valeurs marquées appropriés au contexte (entité, association, attribut…) sont ainsi proposés à l'étudiant.

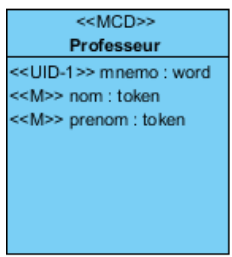

- Une classe UML stéréotypée «MCD» représente une entité.
- Le stéréotype «UID-i» indique un attribut constitutif de l'identifiant naturel i.
- Les types de données des attributs sont repris du W3C et sont inclus dans le profil évoqué.

Figure 3 - Modélisation d'une entité

Nous allons nous appuyer sur un modèle de gestion d'examens pour découvrir quelques aspects de notre formalisme étendu, en particulier la notion de pseudo entité associative qui fait porter des attributs à une association de degré 1:n.

Il est certainement important, à ce stade, de bien montrer à l'étudiant que règle métier et spécification formelle sont deux choses différentes; pour ce faire, nous allons nous appuyer sur le Tableau 1.

|     | Règle métier | Spécification formelle |
| --- | --- | --- |
| ① | Lors de l'enregistrement d'un examen, le temps prévu pour le diriger doit être saisi. | Attribut **dirigeTempsPrevu** obligatoire «M» |
| ② | Un professeur est désigné pour diriger un examen; cette information ne doit pas obligatoirement être saisie lors de la création de l'examen. | Cardinalité 0..1 du rôle **dirige** |
| ③ | Le temps passé par le professeur pour diriger un examen ne peut être saisi que si le professeur est désigné. | Attribut **dirigeTempsPrevu** au sein de la pseudo entité associative **diriger** |
| ④ | Le mnémonique d'un professeur ne doit pas contenir d'espace<br>Le nom d'un professeur ne doit pas avoir d'espaces contigus | Type **word** de l'attribut **mnemo**<br>Type **token** de l'attribut **nom** |
| ⑤ | Un examen est propre à une matière et est identifié par ladite matière et la date de déroulement | Association identifiante de composition «PK» entre **Examen** et **Matiere**. |
| ⑥ | La date de création de l'examen doit être enregistrée automatiquement et ne doit pas pouvoir être modifiée | Attribut **dateCreation** alimenté par la date système *now()* et gelé *{frozen}* |
| ... | | |

Tableau 1 - Correspondance entre règles métier et spécifications formelles

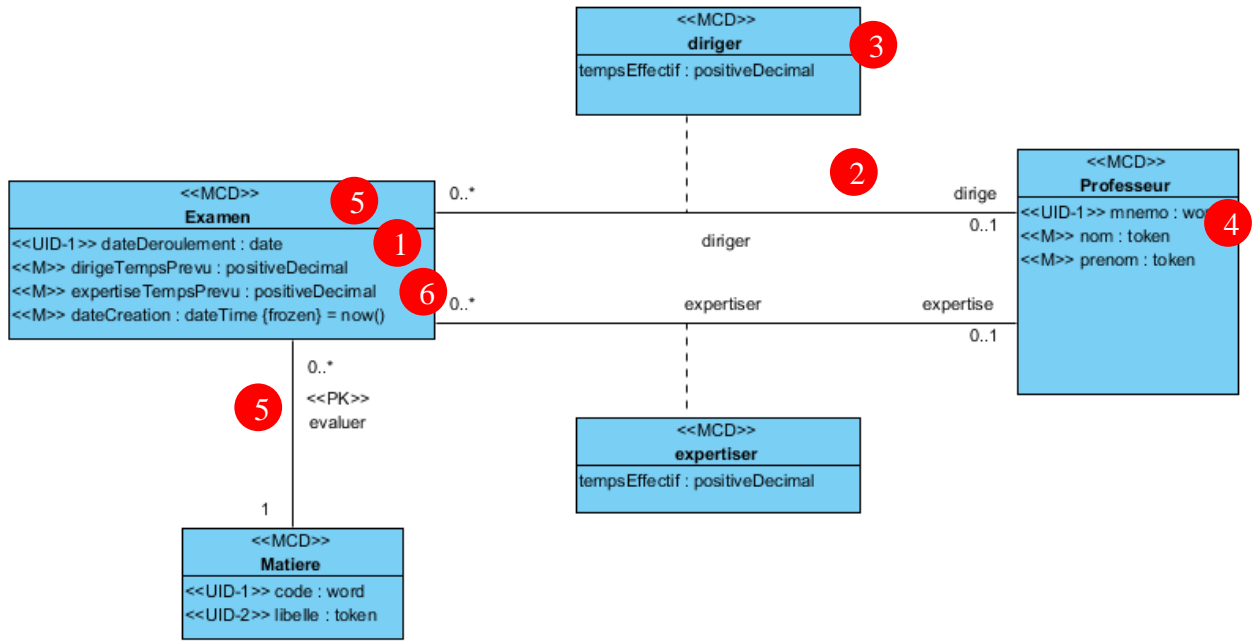

Figure 4 - MCD de gestion d'examens

## 4.2 Contrôle de conformité

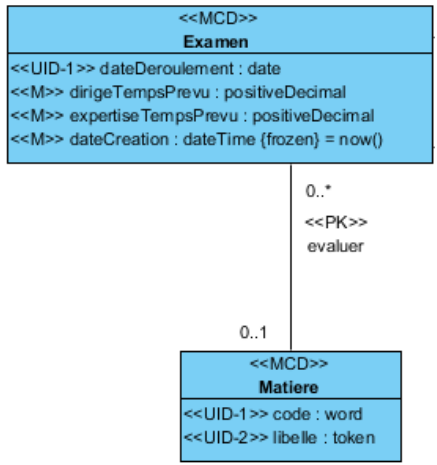

Figure 5 - Erreur de sémantique

Le profil UML du projet MVC-CD guide l'étudiant dans sa tâche de modélisation mais, il ne peut garantir à lui seul que le modèle soit sémantiquement correct. L'automate de transformation offre à l'étudiant une fonctionnalité de contrôle de conformité.

Ci-contre, l'étudiant a fait une erreur sémantique ①; en l'occurrence, une association identifiante de composition doit obligatoirement référer une entité parent.

Le contrôle de conformité va informer l'étudiant des non-conformités de son modèle et lui indiquer leur nature précise; dans le cas présent, que l'extrémité parent d'une association identifiante de composition doit obligatoirement porter une cardinalité 1..1.

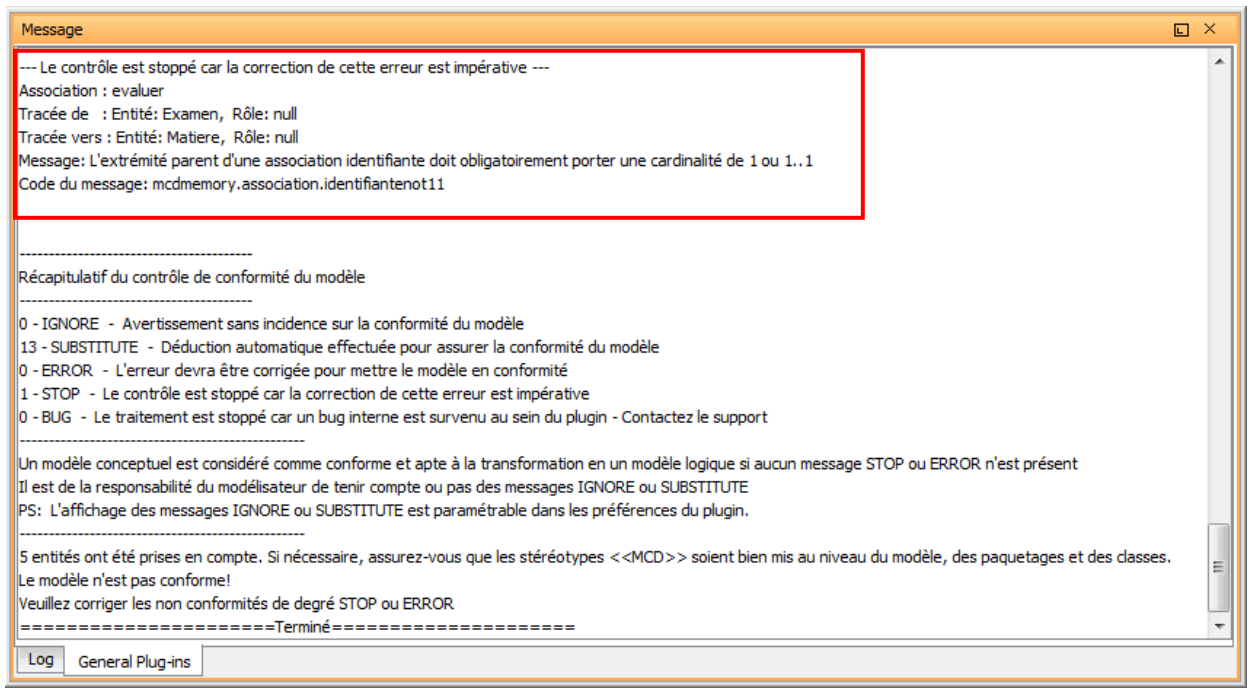

Figure 6 - Fenêtre de quittance du contrôle de conformité

## 4.3 Transformation du MCD

Si le MCD est conforme, l'étudiant peut le transformer en un modèle logique relationnel, relativement indépendant d'un constructeur et ensuite en un modèle physique spécifique à un constructeur.

### 4.3.1 MLD/MPD

Le modèle logique relationnel (MLD) généré est relativement conforme à ce que font de nombreux outils de transformation de famille MDA ou autres.

De manière particulière à notre automate, l'étudiant pourra constater la disparition des pseudos entités associatives et la transformation de leurs attributs en simples colonnes ① dans la table enfant des 2 associations.

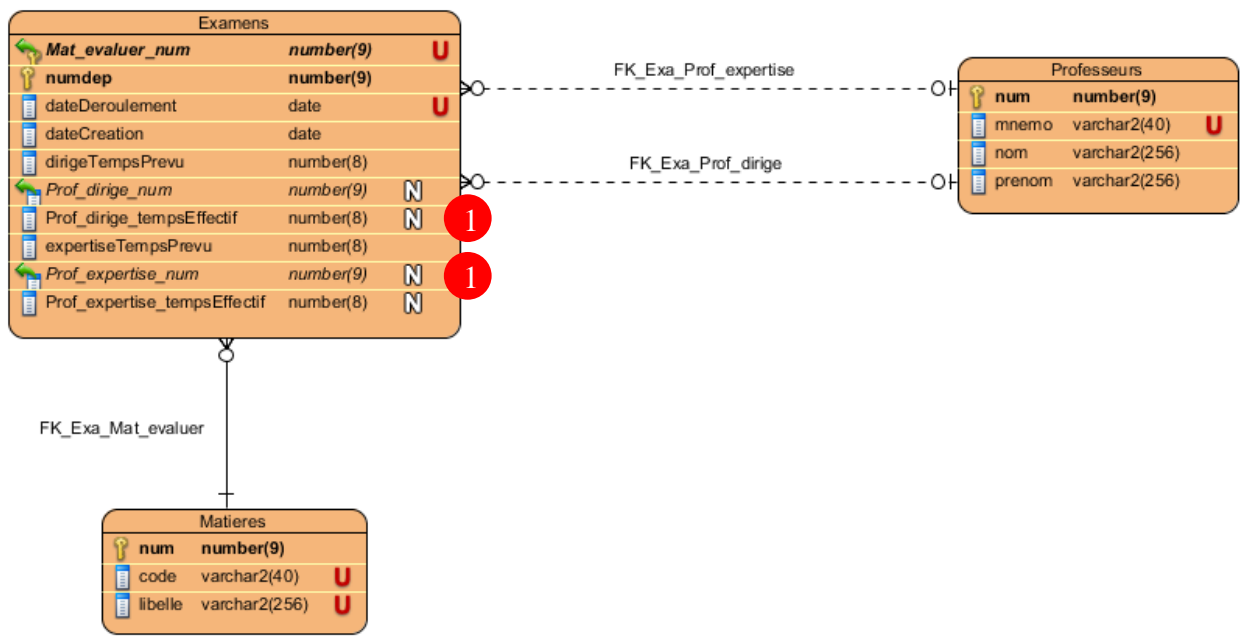

**Figure 7 - MLD de gestion des examens**

### 4.3.2 MPD

L'étudiant attentif aura peut-être relevé le fait que certaines spécifications formelles ne sont, à priori, pas prises en compte par le MLD comme, par exemple, la règle de gestion qui veut que la date de création de l'examen ne soit pas modifiable.

Pour mettre en place une telle règle, l'étudiant a plusieurs choix:

- Il peut la coder au niveau applicatif mais, il faut la recoder pour chaque application qui manipule la table **Examens**. De plus, cette règle et par là l'intégrité des données peut être violée en passant directement par des outils d'administration ou par des connexions offrant la table **Examens** en écriture.

- Il peut écrire du code au sein de la base de données et l'associer à un déclencheur de modification; dans ce cas, il évite les problèmes du codage au niveau applicatif mais, à chaque changement de spécification, il faudra revoir le code et le déclencheur, cela donc au détriment de l'agilité.
- Il peut se dire: "Cette non modification est spécifiée dans le MCD, l'automate ne peut-il pas me générer le code et le trigger associé?".

Oui, l'automate est capable de transformer de nombreuses spécifications que le langage SQL-DDL ne prend en pas charge grâce à une stratégie d'APIs de tables [Note de bas de page 2].

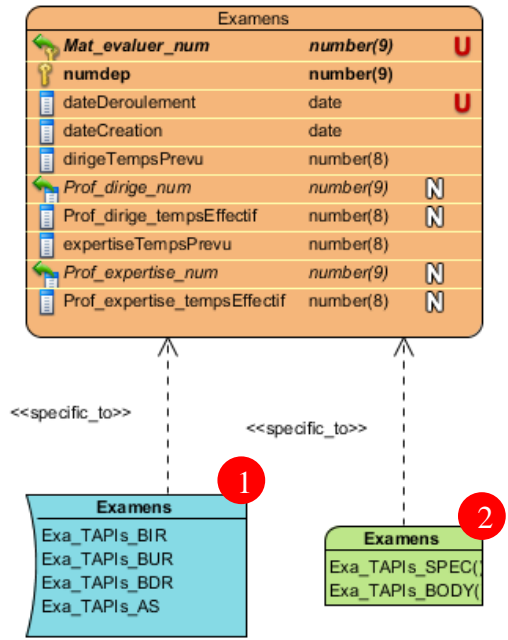

La Figure ci-contre représente le modèle physique des APIs de la table **Examens**. Nous voyons:

① le conteneur des déclencheurs;

② le conteneur des procédures stockées.

Le couple déclencheurs et procédures stockées constitue la base des APIs de tables pour tous les systèmes de gestion de bases de données relationnelles mais, les détails de réalisation sont spécifiques aux divers constructeurs ou produits.

**Figure 8 - MPD de la table Examens**

Pour notre cas de non-modification de la date de création de l'examen, c'est le déclencheur **Exa_TAPIs_BUR** qui sera déclenché lors de la modification d'un tuple (BUR pour Before Update for each Row).

```
TRIGGER Exa_TAPIs_BUR
BEFORE UPDATE ON Examens FOR EACH ROW
    DECLARE
        vl_newrec Examens%ROWTYPE;
        vl_oldrec Examens%ROWTYPE;

    BEGIN

        vl_newrec.Mat_evaluer_num := :NEW.Mat_evaluer_num;
        vl_newrec.numdep := :NEW.numdep;
        vl_newrec.dateCreation := :NEW.dateCreation;       ①
        ...

        vl_oldrec.Mat_evaluer_num := :OLD.Mat_evaluer_num;
        vl_oldrec.numdep := :OLD.numdep;
        vl_oldrec.dateCreation := :OLD.dateCreation;       ②
        ...

        Exa_TAPIs.autogen_column_upd(vl_newrec);
        Exa_TAPIs.autogen_column(vl_newrec);
        Exa_TAPIs.checktype_column(vl_newrec);
        Exa_TAPIs.uppercase_column(vl_newrec);
        Exa_TAPIs.column_PEA(vl_newrec);
        Exa_TAPIs.frozen_column(vl_newrec, vl_oldrec);     ③

        :NEW.Mat_evaluer_num := vl_newrec.Mat_evaluer_num;
        :NEW.numdep := vl_newrec.numdep;
        :NEW.dateCreation := vl_newrec.dateCreation;       ④
    END;
```

**Figure 9 - Extrait de code du déclencheur sur modification de la table Examens pour Oracle**

Que fait le déclencheur **Exa_TAPIs_BUR**?

① Il met les données telles qu'elles seront après modification dans une variable.

② Il fait de même pour les données telles qu'elles étaient avant modification.

③ Il appelle les procédures de traitement; dans notre cas, il s'agit de la procédure **frozen_column()** en lui passant les deux variables décrites précédemment.

④ Il modifie les données pour prendre en compte les éventuels changements apportés par les procédures stockées.

```
PROCEDURE frozen_column(    pio_newrec IN OUT Examens%ROWTYPE,
                            pio_oldrec IN OUT Examens%ROWTYPE) IS
BEGIN
  IF pio_newrec.dateCreation <> pio_oldrec.dateCreation THEN
    raise_application_error(-20001, 'Table: Examens , Colonne: dateCreation ,
                          mpd.constraint.mess.err.column.frozen ,
                          La valeur de la colonne n''est pas modifiable');
  END IF;
END;
```

Figure 10 - Code de la procédure de contrôle de non-modification pour Oracle

Que fait la procédure **frozen_column**()?

- Elle assure le respect de notre règle métier initiale, à savoir que la date de création de l'examen ne doit jamais être modifiée, en provoquant une erreur si la valeur après modification est différente de la valeur avant modification. Naturellement, cette erreur d'intégrité des données devra être remontée explicitement par la couche applicative.

## 4.4 Exigences non fonctionnelles

Certaines exigences exprimées vis-à-vis du système d'information informatisé ne relèvent pas d'une fonctionnalité particulière mais de contraintes qui peuvent s'appliquer à tout ou partie d'un modèle et cela, en certains cas, indépendamment de la signification particulière du modèle.

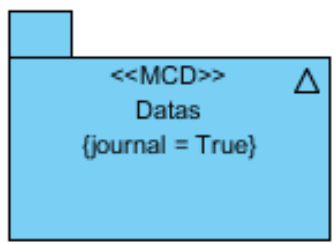

Figure 11 - Journalisation de toutes les entités d'un modèle

Pour notre exemple et dans la perspective de la nécessité de traçabilité requise par les systèmes qualité, il peut être nécessaire de garantir la traçabilité de toutes les opérations d'ajout, de modification ou encore de suppression faites au sein des données de notre gestion des examens.

Pour garantir la traçabilité des opérations, l'étudiant, guidé par le profil MVC-CD, spécifie que tout le modèle [Figure 11], une partie du modèle ou une entité particulière doit être journalisé.

Comment notre automate met-il en place la journalisation? Comme Oracle le faisait avec Designer, il s'appuie sur les APIs de tables et, pour ce faire:

- ① crée, pour chaque entité à journaliser, une table de journalisation qui contiendra un tuple pour chaque opération effectuée dans la table à journaliser;
- ② ajoute dans les déclencheurs de la table à journaliser un ordre de journalisation pour les opérations d'ajout, de modification et de suppression;
- ③ crée la procédure de journalisation proprement dite.

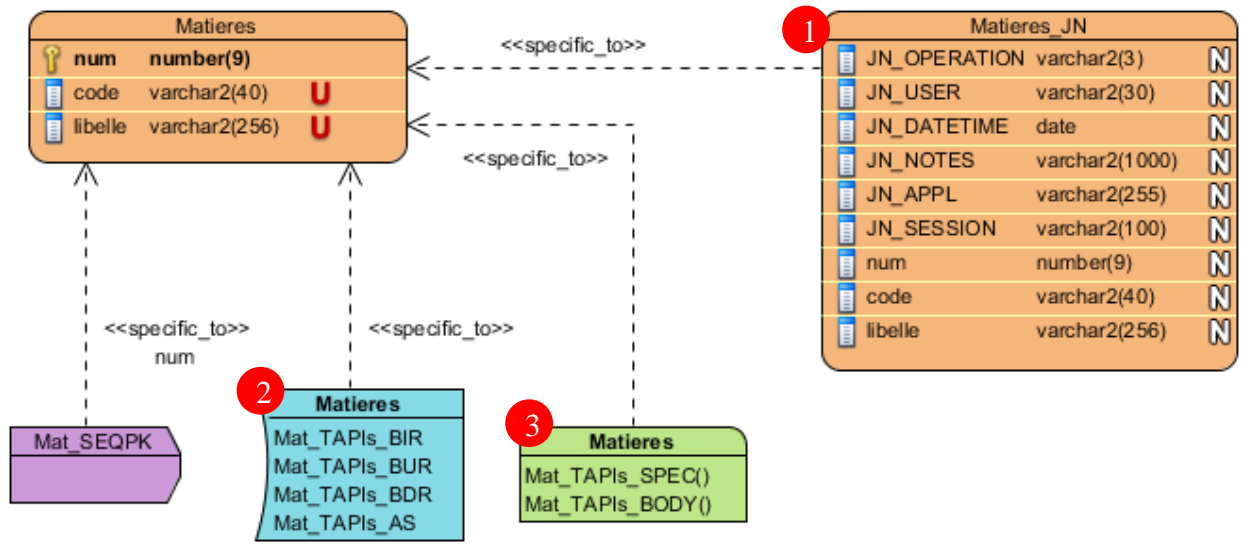

Figure 12 - Journalisation de la table Matieress

Après déploiement du modèle de données et réalisation de commandes d'insertion, de modification et de suppression, l'étudiant peut constater l'alimentation automatique des tables de journalisation faite à partir des APIs de tables.

| JN_OPERATION | JN_USER | JN_DATETIME |  |  |  | NUM | CODE | LIBELLE |
|---|---|---|---|---|---|---|---|---|
| INS | A | 02.03.15 | .. | .. | .. | 203 | CH | Chimie |
| UPD | B | 05.03.15 | .. | .. | .. | 203 | CH | Chimie et physique |
| UPD | A | 08.03.15 | .. | .. | .. | 203 | CHP | Chimie et physique |
| DEL | A | 13.03.15 | .. | .. | .. | 203 | (null) | (null) |

Figure 13 - Vue de la trace des opérations effectuées sur l'enregistrement 203 de la table Matieres

## 4.5 Qu'en est-il de l'agilité?

A ce stade, l'étudiant a probablement été sensibilisé aux gains de productivité que procure une démarche d'ingénierie pilotée par les modèles adossée à des outils de transformation et de génération de code. Le développement peut être qualifié d'agile en ce sens que l'impact d'un changement de règles métier sur le code assurant l'intégrité des données est minimisé grâce à l'automatisme des APIs de tables. Mais qu'en est-il de l'agilité du modèle lui-même?

Pour aborder cette question, revenons au MCD de la Figure 4; l'association identifiante de composition entre **Matiere** et **Examen** est transformée au niveau logique en une relation identifiante [Figure 7]. Si cette relation identifiante, matérialisée par la colonne de clé étrangère **Mat_evaluer_num** de la table **Matieres,** devait ne plus être identifiante ou ne plus exister, la contrainte de clé primaire devrait être modifiée; mais, la modification d'une contrainte de clé primaire ne peut pas se faire, ou à tout le moins difficilement et périlleusement, pour un système d'information informatisé opérationnel. A priori donc, le modèle logique relationnel, face à certains changements, est bloquant et pourrait être une entrave à l'agilité des systèmes d'information.

Pour éviter ce blocage, il faudrait renoncer aux relations identifiantes mais alors, nous perdrions les contraintes d'intégrité implicites qu'elles supportent. Dans l'optique où le concepteur de système d'information ne veut pas recourir aux relations identifiantes pour ne pas mettre en péril la faculté d'agilité de son système d'information, l'automate propose un mode de transformation alternatif. Ce mode de transformation alternatif ne crée que des relations non identifiantes et les contraintes implicites qu'auraient supportées ces relations sont remplacées par des contraintes explicites.

Nous voyons ci-dessous que la relation entre **Examens** et **Matieres** est devenue non identifiante et en corolaire la colonne de clé étrangère **Mat_evaluer_num** ne fait plus partie de la contrainte de clé primaire.

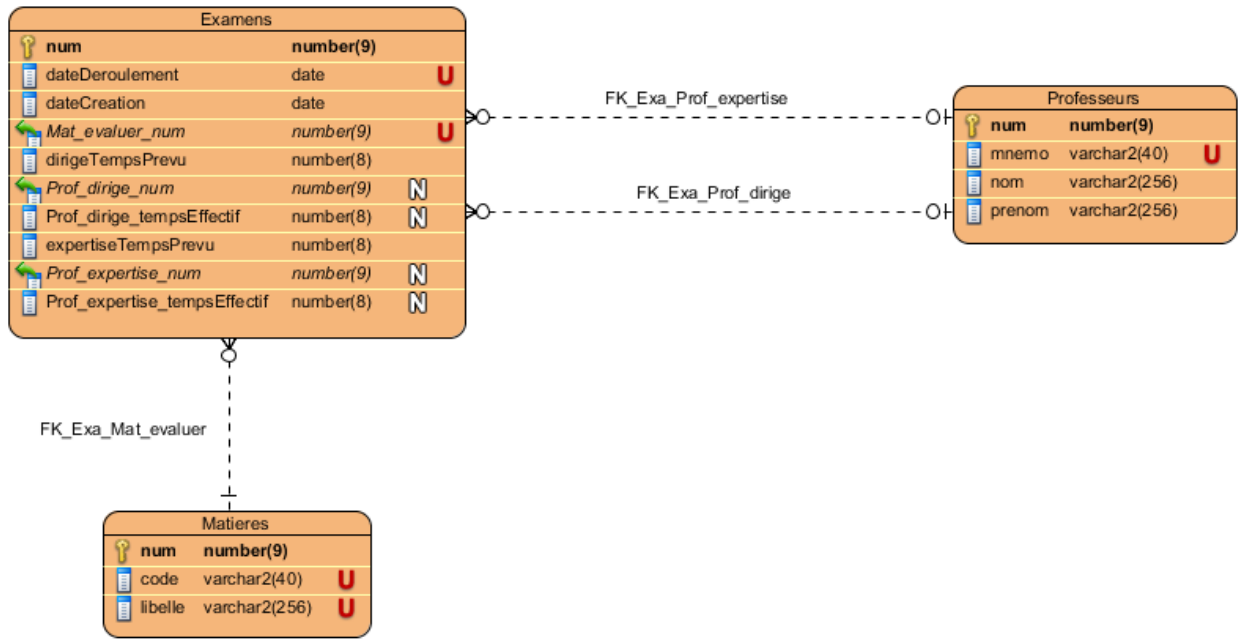

Figure 14 - MLD de gestion d'examens basé exclusivement sur des relations non identifiantes

Par ailleurs, l'automate a ajouté, dans les spécifications de la table **Matieres,** la contrainte explicite ci-dessous qui va assurer la règle métier ⑤ du Tableau 1.

| Name | Type | Constraint |
|---|---|---|
| UID1_Exa | Unique | UNIQUE (Mat_evaluer_num, dateDeroulement) |

Figure 15 - Contrainte d'intégrité explicite

# 5   Conclusion

En introduction, nous nous posions la question "Comment réconcilier modèles, dans l'esprit actuel de MDA et génération de code, dans la ligne de ce que faisaient les outils CASE ?". Nous nous posions cette question dans la perspective d'augmenter l'agilité des logiciels de gestion et, par là, diminuer les contraintes que peuvent exercer les systèmes d'information informatisés sur les organisations.

Nous avons concentré notre recherche sur la modélisation des données en travaillant sur deux axes principaux:

- Élaboration d'une extension du formalisme usuel de modélisation.
- Conception et développement d'un automate de transformation.

Nous avons pu montrer, à nos étudiants, au travers de nos cours de conception de SII, que le recours à des APIs de tables générées à partir de spécifications formelles étendues rend plus agile le développement de logiciels de gestion; en effet, de nombreuses contraintes d'intégrité de données ne sont plus à coder manuellement. De plus, le recours à des transformations de modèles basées sur des mécanismes découplant structure et contraintes offre des opportunités d'agilité non pas seulement au niveau du code mais aussi au niveau de l'analyse et de la modélisation.

Nous constatons que nos étudiants sont plus attentifs à la nécessité de traduire avec précision les règles métiers du système d'information informatisé à concevoir en spécifications formelles. Ils sont toujours très intéressés par le devenir de ces spécifications formelles; les travaux pratiques qu'ils réalisent avec l'automate de transformation les confortent à modéliser avec soin. De plus, nous remarquons que certains étudiants utilisent l'automate durant la suite de leur formation ce qui tend à montrer l'intérêt qu'ils trouvent à une démarche d'ingénierie pilotée par les modèles.

Par ailleurs, nous remarquons que les étudiants de première année, qui n'ont souvent aucune expérience en modélisation des données, semblent convertir les règles métier (des données d'exercices textuelles) en spécifications formelles de manière "naturelle". En outre, la fonctionnalité de contrôle de conformité de l'automate MVC-CD leur permet d'effectuer une auto correction syntaxique de leurs modèles; il est plaisant de voir les étudiants comprendre et corriger par eux-mêmes leurs erreurs de formalisme.

En terme pédagogique, il nous semble intéressant de poursuivre le travail initié en prenant en compte:

- d'autres technologies de persistance de données (XML, orienté objet, classeurs de feuilles de calculs…) car beaucoup de nos étudiants en informatique de gestion ne pensent, ne voient ou n'imaginent que le modèle relationnel pour assurer la persistance;
- la spécification d'interfaces utilisateurs et leur génération pour montrer aux étudiants la dépendance implicite entre structure de données et interfaces utilisateurs;
- OCL pour spécifier des contraintes qui ne peuvent l'être de manière déclarative;
- le portage du formalisme étendu et de l'automate de transformation dans un environnement ouvert.

# 6 Bibliographie

# Annexe A - Méthodologie d'élaboration d'une extension du formalisme usuel de modélisation des données

Les outils CASE des années 1970-2000 étaient souvent propriétaires et s'appuyaient sur des modèles propriétaires ou à tout le moins peu ou pas partagés. Ce manque d'ouverture des outils CASE a probablement été aussi un frein[7] à leur propagation; dès lors, notre premier choix méthodologique a été de retenir le standard ouvert de l'OMG, UML, comme langage de modélisation.

Les modèles conceptuels de données au sens de MCD de Merise, de ER Model de P. Chen ou encore de Modèle du domaine de UP véhiculent des concepts éprouvés et utilisés depuis près de 40 ans pour certains; il nous a paru évident de bâtir notre démarche sur lesdits concepts.

Au cours de notre expérience de modélisation des données, que ce soit sous forme de mandats pour des entreprises tierces, de travaux de recherche ou d'enseignement, nous avons eu l'occasion de travailler avec :

- les règles de construction d'un modèle conceptuel de données (MCD) de Merise (Tardieu et al, 1986);
- les règles de construction d'un modèle ER Model de P. Chen;
- les règles de construction d'un modèle de classes UML (Rumbaugh et al., 2004), et le mapping à faire pour représenter un modèle conceptuel de données (Kettani et a., 1999);
- les règles de construction d'un modèle du domaine de UP (Jacobson et al., 1999) et (Larman, 2005);
- les règles de construction d'un modèle ER Model mises en œuvre par l'outil CASE Pro*Kit Workbench de McDonnel Douglas;
- les règles de construction d'un modèle ER Model mises en œuvre par Oracle dans l'outil CASE Designer (Koletzke et Dorsey, 1999).

Au fil du temps, nous avons eu l'occasion de critiquer, par la mise en pratique en de multiples situations, ces différentes règles. Et, sans que cela fut un travail identifié en tant que tel, nous avons réalisé une synthèse de ces différentes règles ; progressivement, nous en avons conforté certaines, modifié ou rejeté d'autres ou encore créé de nouvelles.

Pour cette recherche, nous avons recensé les règles de construction de modèles conceptuels de données qui nous semblent pérennes, les avons classées, les avons discutées et en finalité, nous en avons fait une typologie. Ensuite, pour chaque type de règle, nous avons défini une manière de la représenter, sous forme de spécification formelle, en nous appuyant sur les opportunités offertes par UML ; nous nous sommes fixé comme modalité de représentation que:

- les spécifications formelles soient visibles dans la représentation graphique du modèle. (Rumbaugh et al., 2004) "Les éléments de présentation véhiculent l'aspect visuel… Ils n'ajoutent pas de signification mais ils organisent la présentation du modèle afin que celui-ci puisse être exploité."
- les spécifications doivent pouvoir être posées sans nécessiter d'anticiper une transformation à venir ; nos spécifications conceptuelles doivent être intelligibles en tant que telles et indépendantes de tout choix d'implantation (Base de données relationnelle ou orientée objet ou XML ou encore un classeur de feuilles de calcul ou de simples fichiers).

---

[7] En plus des raisons invoquées dans la revue de littérature (Chapitre 2.2)

Chaque fois que cela était possible, nous avons modélisé les types de règles dans un profil UML. Un profil UML est un méta modèle et notre profil est le méta modèle de réalisation d'un MCD conforme à nos règles de modélisation.

En plus de cette démarche de recensement des règles de construction de modèle conceptuel dans la littérature et dans les souvenirs de nos expériences passées, nous avons aussi fait la démarche inverse; nous sommes allés rechercher, dans le code de gestion de l'intégrité des données, le code qui est répétitif et qui pourrait être généré à partir de spécifications formelles. Cette recherche dans le code n'est pas le fait de ce travail de recherche particulier mais le résultat de l'observation de multiples développements passés où nous avons le sentiment de réinventer la roue. Par contre, dans ce travail de recherche, nous avons identifié, classé et analysé de multiples situations de code répétitif ; de cette analyse, de code répétitif, nous avons dégagé, lorsque cela nous était possible, des spécifications formelles. Nous avons intégré ces spécifications formelles dans notre typologie et les avons modélisées au sein du profil UML.

## Annexe B - Méthodologie de conception et développement d'un automate de transformation

Nous avions choisi, il y a quelques années, l'outil de modélisation Visual Paradigm for UML[8] (VP) pour notre enseignement et nos travaux de recherche.

Dans le cadre de notre recherche, VP nous permet de:

- modéliser les données, de niveau conceptuel, en respectant le formalisme étendu que nous visons;
- modéliser un profil de MCD conforme (moyennant quelques petites restrictions sans incidence dommageable);
- générer les scripts SQL-DDL à partir d'un modèle logique/physique de données ;
- générer les triggers et procédures stockées à partir d'un modèle physique de données.

VP dispose d'un automate de transformation de MCD en MLD. Cet automate, que nous avions évalué en version 9, est très basique ; à titre d'exemple, nous avons constaté qu'il n'est pas capable de transformer une entité associative. VP offre la possibilité de réaliser des "plugins" qui interagissent avec son référentiel.

---
[8] http://www.visual-paradigm.com/

À partir des éléments ci-dessus, nous avons décidé d'utiliser VP pour ses capacités de modélisation et de génération de code. Pour la transformation d'un MCD en un MLD/MPD relationnel, sans perte de sémantique par rapport au formalisme étendu que nous avons défini, nous allons réaliser, nous-mêmes, un automate au sein de VP. Nous avons nommé cet automate MVC-CD.

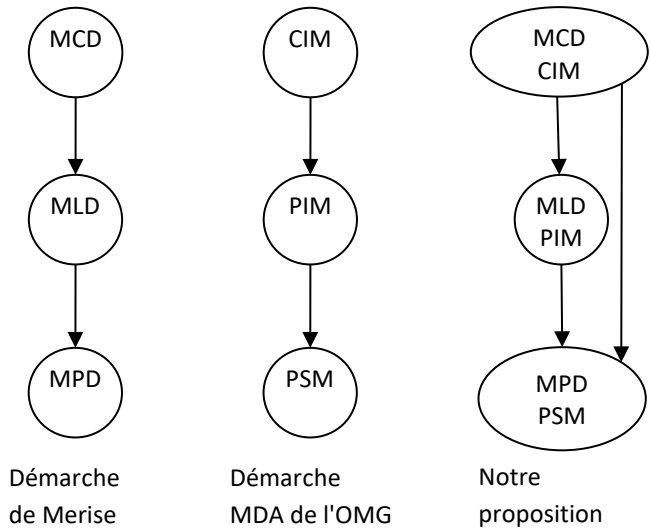

Figure 16 - Niveaux d'abstraction des modèles

La pratique courante de transformation d'un modèle conceptuel de données est de passer par un modèle logique, orienté technologie, et ensuite d'arriver à un modèle physique, orienté produit. Cette démarche est intéressante, car elle permet d'enrichir un modèle logique avant de le transformer en un modèle physique.

Cet enrichissement a tout son sens lorsque l'enrichissement porte sur des spécificités technologiques ; par contre, si l'enrichissement découle de règles qui sont connues et formalisables au niveau conceptuel il nous faudrait le faire à ce niveau pour ne pas perdre ou oublier ces règles.

La transformation automatisée est un élément important de réduction des coûts et délais de développement des logiciels mais elle n'améliore la capacité d'agilité que si elle est itérative et incrémentale. Pour réaliser cette transformation itérative et incrémentale, nous nous sommes basés, pour l'essentiel, sur la méthodologie mise en œuvre par l'ancien outil CASE Oracle Designer et parfaitement décrite par (Koletzke et Dorsey, 1999) au chapitre 12.

Par rapport à l'approche d'Oracle Designer, nous allons apporter les deux modifications suivantes:

- Remplacement de l'enrichissement de niveau logique (MLD ou PIM) au profit d'un enrichissement directement au niveau conceptuel (MCD ou CIM) comme indiqué ci-dessus [Figure 16].
- Plus de souplesse au niveau de la prise en compte des contraintes induites par des données déjà existantes dans un processus itératif et incrémental. Lorsque des données sont déjà existantes, Oracle Designer offre au modélisateur le choix d'effectuer un changement ou pas entre deux itérations. Ce choix binaire nous le modulons en offrant la possibilité de faire un changement forcé (implique une altération potentielle de données existantes; le cas d'une diminution de la taille d'une donnée) ou un changement préservé (le cas d'une augmentation de la taille d'une donnée).

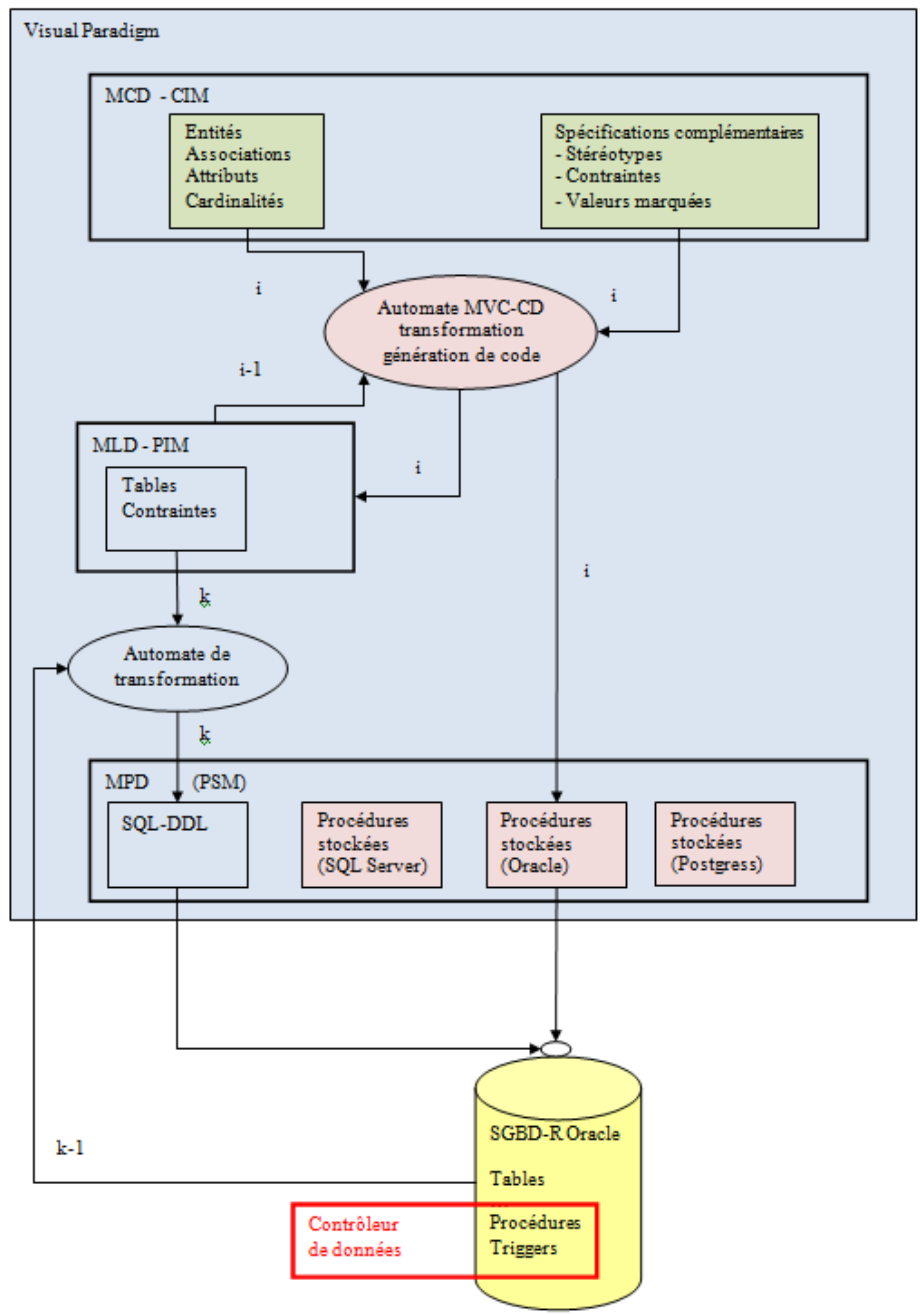

**Figure 17 - Représentation symbolique de l'automate MVC-CD et son environnement**

Dans l'optique de démontrer la faisabilité de notre proposition d'automatisation, nous nous sommes limités à réaliser les modèles physique MPD-PSM pour Oracle; toutefois, nous avons d'ores et déjà conçu l'architecture de l'automate pour prendre en compte d'autres constructeurs ou produits de bases de données relationnelles.